# Post-Route Alleviation of Dense Meander Segments in High-Performance Printed Circuit Boards

Tsun-Ming Tseng[†]  Bing Li[†]  Tsung-Yi Ho[‡]  Ulf Schlichtmann[†]
tsun-ming.tseng@tum.de  b.li@tum.de  tyho@csie.ncku.edu.tw  ulf.schlichtmann@tum.de
[†]Technische Universitaet Muenchen       [‡]National Cheng Kung University
Arcisstrasse 21, 80333 Munich, Germany       No. 1, University Rd., Tainan City, Taiwan

*Abstract*—Length-matching is an important technique to balance delays of bus signals in high-performance PCB routing. Existing routers, however, may generate dense meander segments with small distance. Signals propagating across these meander segments exhibit a speedup effect due to crosstalks between the segments of the same wire, thus leading to mismatch of arrival times even with the same physical wire length. In this paper, we propose a post-processing method to enlarge the width and the distance of meander segments and distribute them more evenly on the board so that the crosstalks can be reduced. In the proposed framework, we model the sharing combinations of available routing areas after removing dense meander segments from the initial routing, as well as the generation of relaxed meander segments and their groups in subareas. Thereafter, this model is transformed into an ILP problem and solved efficiently. Experimental results show that the proposed method can extend the width and the distance of meander segments about two times even under very tight area constraints, so that the crosstalks and thus the speedup effect can be alleviated effectively in high-performance PCB designs.

## I. Introduction

In high-performance printed circuit boards (PCBs), delay matching between bus signals has become a mainstream problem which is considered in modern PCB routers [1]–[4]. In [1] the delay matching problem is solved by using a Lagrangian model to allocate resources for wire snaking. In [2] this problem is solved with the help of slant symmetric grids. The method in [3] transforms this length matching task to an area assignment problem and proposes a gridless framework using bounded-sliceline grids. The method in [4] successfully routes given designs considering matching wire lengths and wire shapes, while still keeping high efficiency in using routing resources. In these methods, wires which do not have sufficient lengths are extended by creating snaking patterns in the routing, on the assumption that signals across wires with the same length have the same delay. These patterns have a high routing density and can be relatively easily modeled, and therefore have gained wide acceptance. Fig. 1 illustrates such a snaking pattern, which is henceforth called *meander segment*, and a wire with concatenated meander segments. NE and FE in Fig. 1 are abbreviations of near end and far end, respectively.

Accompanying the adoption of the meander segments as a delay compensation method, much research effort has been put into the study of their delay characteristics [5]–[8]. In [5] it is shown that crosstalk noise between meander segments has an accumulation effect in high-speed designs and a *speedup* effect on the wires may appear in such patterns of high density. In [6] a moment technique is proposed to approximate the delays of wires with meander segments, and in [7] a method with finite-difference timing domain is used for the qualitative prediction of such delay lines. Furthermore, in [8] a linear model is formulated to control the wire delay by adjusting the number of meander segments on a fixed-shape wire.

According to [5]–[8], when a signal travels across meander segments, the crosstalks between the segments of the same wire accumulate gradually. Therefore, the signal can reach the sinking pin earlier than predicted by the wire length, thus causing delay mismatch between bus signals. Consider the pattern in Fig. 1 with nine wire segments. At time zero, the main signal switches at the near end of wire 1 and propagates from bottom to top. This signal stimulates crosstalk signals at the near ends of the other wire segments. Assume that the total propagation time of the signal from the near end of one wire segment to the far end of the next wire segment, or from the far end to the other near end, is $t_d$. At time $t_d$, the main signal reaches the far end of wire 2 and stimulates a new crosstalk voltage. This new voltage superposes on the crosstalk signal triggered by the main signal at time zero, which reaches the far end of wire 3 also at $t_d$. This superposition process continues when the main signal propagates across each wire segment, and finally the crosstalk voltage may surpass the threshold of logic switching before the main signal, thus leading to a speedup effect on the wire.

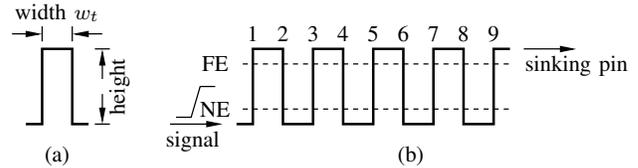

Fig. 1: (a) Meander segment (b) A wire with concatenated meander segments.

Despite the crosstalk noise between meander segments, they are still widely used in floorplan-like or area-assignment-like routing methods such as [1]–[4], because they can be applied relatively easily to match wire lengths and can be adjusted flexibly. Other matching patterns, for instance, the flat spiral delay line [9] or the concentric delay line [7], impose more computational complexity and are still not widely applied, especially in existing commercial tools.

The major contributions of this paper include a mathematical model for meander patterns in a subarea and space sharing by multiple wire groups, as well as an iterative algorithm to find the largest possible width for the newly created meander segments, which are used to compensate the wire lengths after we remove the dense meander segments from the original routing. The resulting routing has a similar shape and the same wire lengths as in the original routing, but with meander segments of enlarged width. After applying the proposed method, wire delays can be estimated by wire lengths, so that the accuracy of delay matching can be improved. In addition, the proposed method adjusts routing results from other routers, which have already determined the basic routing patterns and wire lengths, so that it can be integrated into an existing PCB design flow seamlessly.

The rest of the paper is organized as follows. In Section II we give the formulation of the problem to be solved in this paper. In Section III we explain the proposed method to alleviate dense meander segments in details. We then show results with several PCB designs in Section IV and conclude our paper in Section V.

## II. Problem Formulation

Different methods can be deployed to handle the dense meander segments in high-performance PCBs to avoid or alleviate the speedup effect and thus the delay mismatch on bus signals. An intuitive method is to extend the lengths of the wires to match their delays. But this direct solution relies on the prerequisite that wire delays can be evaluated accurately, which is not an easy task in view of different numbers and varying shapes of meander segments. Another method is to form meander segments with different heights so that the crosstalk signals are not aligned and can not accumulate. But this method needs a close look on the signal propagation and is very sensitive to post-routing changes.

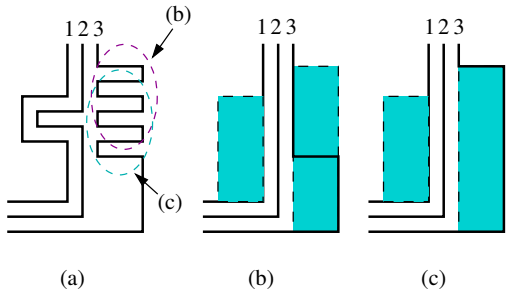

Fig. 2: Removal of dense meander segments.

In this paper, we try to alleviate the speedup effect by increasing the width of meander segments, that is, the distance between the wire segments forming such a pattern. The basic idea of this method is to remove the dense meander segments from the original routing and use the free space to form new patterns for wire length compensation. Instead of simply enlarging the widths of existing meander segments in the given routing, which is rarely optimal when multiple wires compete for the available free spaces, we establish a mathematical model for meander segments sharing the same free space and the routing patterns in subareas. Thereafter, this model is transformed into an integer linear programming (ILP) problem and processed by a solver to provide global guidance for generating new relaxed meander segments. An iterative algorithm is also applied to find the largest possible width for the meander segments.

The inputs of the proposed method include the original PCB routing and the given area constraints by which wire segments generated by the post-processing method are confined. The output is a refined PCB routing and the objective of the proposed method is to enlarge the widths of the dense meander segments as much as possible without changing the original wire lengths or violating the given area constraints.

## III. ALLEVIATION OF MEANDER SEGMENTS

In this section we explain the proposed method to enlarge the width of meander segments in a given routing. The proposed framework includes a mathematical model for patterns in a subarea and space sharing by multiple wire groups. In addition, an iterative algorithm is used to find the largest possible width $w_t$ for the newly created meander segments, while complying with given area constraints and guaranteeing that the distance between any two wires is larger than $d_m$ required for manufacturing.

### A. Removal of dense meander segments

The first step of the proposed method is to delete dense meander segments with width smaller than a predefined value $w_t$. New meander segments with a minimum width $w_t$ are grown in the space available thereafter. Fig. 2 shows an example of deleting dense meander segments with widths no larger than three units. The shaded areas with dashed boundaries outline the available free spaces for growing new meander segments. In this operation, the dense meander segments on wire 1 and 2 can be removed simply, but the ones on wire 3 can be viewed from different sides of this wire. Viewed from the left side or the right, the meander segments on wire 3 can be deleted differently and the results are shown Fig. 2(b) and (c), respectively.

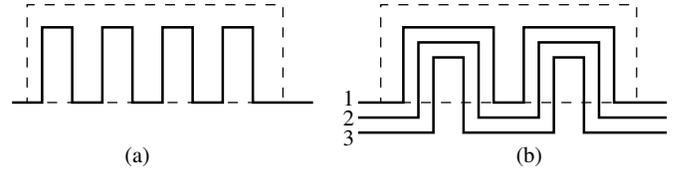

Fig. 3: Growth of meander segments. (a) Single wire. (b) Wire groups.

Although the free space in Fig. 2(b) is split into two parts, we can not assert that Fig. 2(c) is *globally* better than Fig. 2(b), because the relations of wires and all free spaces along the bus determine the best configuration together. In the proposed method, we trace the wires from one direction to delete meander segments. For example, if we trace the wires from top to bottom in Fig. 2(a), we can identify the free space as in Fig. 2(b). Comparing both cases, we notice that the case in Fig. 2(c) can be formed from Fig. 2(b) by shifting upwards the horizontal wire segment that splits the free space. In the proposed method, we first try to grow new dense meander segments in the available free space created by one-directional tracing. If some wires need more space after the first iterations, we scan the wire segments on them and shift these segments to form larger spaces.

### B. Growth of meander segments and space sharing

In this section we explain the modeling of meander segments in a given free space and the sharing of a free space by multiple wire groups. The generated constraints will be used in Section III-C to find an optimal solution.

#### 1) Modeling the patterns in a given free space

In a given free space the growth of meander segments with minimum width $w_t$ on *one* wire can be performed relatively easily by calculating the allowed number of meander segments in this area. Fig. 3(a) shows such an example. In the routing of a bus, however, more than one wire can exist at a side of a free space. For example, three wires at the bottom of the free space in Fig. 3(b) can form meander segments in the free space at the same time. With this pattern, wires which are below the other wires still have a chance to use the free space to extend their lengths, but at the expense that the widths of the meander segments of the upper wires should be increased, so that fewer meander segments and thus shorter length growth for the outer wires to compensate the removed wire segments can be created. Comparing Fig. 3(a) and (b), we can observe that this is a tradeoff between sharing the free space with a group of wires and maximizing the length compensation of individual wires. The complexity comes from the fact that it is not easy to determine which wires should be pushed into the free space, and how many meander segments should be formed on each wire. In addition, there exists a dependency between these newly formed meander segments. For example, wire 3 can be pushed into the free space only when wire 1 and 2 are pushed upwards.

To solve the problem described above, we formulate a flexible model to handle the number of meander segments in the free space and the dependency between multiple wires. A general analysis of the new meander segments in a free space is shown in Fig.4, where four upward meander segment groups (msg) are illustrated to show different relations of the wires. In $msg_1$ the lower meander segment has enough height to take a part of the internal space of the upper segment, so that the width of the upper segment must be at least two times of the minimum wire distances $d_m$ larger than the width of the lower meander segment. In $msg_2$ only the left vertical wire segment of the upper meander segment has conflict with the lower meander segment so that the width of the upper meander segment needs only to be increased by one $d_m$. In $msg_3$, no conflict exists so that both meander segments can have the width equal to $w_t$. In the last group the lower wire does not have a meander segment because no wire length compensation is needed. In this case, the width of the upper meander segment is equal to $w_t$. In addition to these segment groups, intermediate upward segments may be grown, for example, the one shown with dashed line in Fig. 4. In modeling all the relations in the meander segment groups, the intuitive model in Fig. 4 needs to assign variables to all the heights of the meander segments, and to describe the possible relations of the wires with many constraints. For example, the widths of meander segments on wire 1 depend on the heights of the ones on wire 2. However, not all the flexibility provided in Fig. 4 is really necessary, because different wire heights may lead to inefficiency in using free spaces, for example, the free space between $msg_1$ and $msg_3$ and above $msg_2$ may be wasted due to the two larger heights of $msg_1$ and $msg_3$. When the model is processed by a solver, such a case is rarely selected if the objective of optimization includes the maximization of wire length compensation and the usage of free spaces.

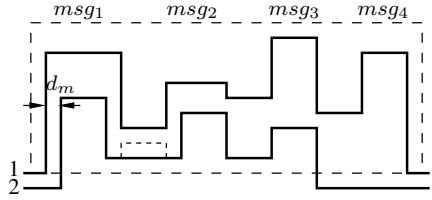

Fig. 4: General model of meander segments.

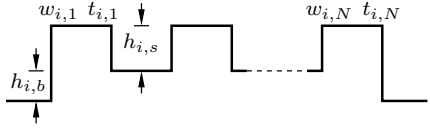

Fig. 5: Simplified meander segment model of a wire.

Based on the general formulation in Fig. 4, we propose a simplified model with fewer variables and constraints. The variables for the $i$th wire are shown in Fig. 5. This model can be considered as two parts, where $N$ *subordinate meander segments* with heights $h_{i,s}$ grow at the top of the *base meander segment,* which has the height $h_{i,b}$ and spans across all the horizontally available space. In this model, the subordinate meander segments increase the length of the $i$th wire, while the base meander segment shifts the free space so that the wires below it also have access to the free space. To model the heights of the pattern in Fig. 5, we need only two variables $h_{i,b}$ and $h_{i,s}$, instead of the $2N$ variables if the general formulation in Fig. 4 is used. This simplified model sacrifices the flexibility in selecting heights of the subordinate meander segments. However, this flexibility does not contribute to the maximization of wire length compensation much because the irregular heights in Fig. 4 may actually waste the free space in growing meander segments as discussed before. In the simplified model, we have also ignored the intermediate meander segments shown with the dashed line in Fig. 4. These possibilities are explored by the iterations in the proposed method in Section III-C. As shown in Fig. 5, the width of the $j$th subordinate meander segment of the $i$th wire is represented using $w_{i,j}, k = 1, \ldots N$. In addition, we assign a 0-1 variable $t_{i,j}$ for each subordinate meander segment in Fig. 5 to model its appearance in the final routing. The usage of these variables will be explained in what follows.

With the simplified model in Fig. 5, we can now model the growth of meander segments from a group of wires into a free space, as shown in Fig. 6, where three wires and two subordinate meander segments on each wire are used as example. Each wire in Fig. 6 has the variables assigned as in Fig. 5. For the $i$th and $(i+1)$th wires, the distance between the horizontal wire segments should always be larger than the minimum wire distance $d_m$ to guarantee manufacturability. These constraints can be described as

$$(h_{i,b} + h_{i,s} + d_m) - (h_{i+1,b} + h_{i+1,s}) \geq d_m \quad (1)$$
$$(h_{i,b} + d_m) - h_{i+1,b} \geq d_m \quad (2)$$

where $i = 1, \ldots M - 1$. $M$ is the number of wires and equal to 3 in Fig. 6. Here we have assumed that the wire distance in the original wire group is $d_m$ to simplify the expressions in (1) and (2). The constraint (1) describes the relations of horizontal segments at the top of the subordinate meander segments, and (2) the relations at the bottom of the subordinate meander segments. From (1)–(2), we can observe that the subordinate meander segments on the first wire have the largest height. In order to fit the grown patterns into the free space, this height should be no larger than the height of the given free space, that is,

$$h_{1,b} + h_{1,s} \leq h \quad (3)$$

where $h$ is the height of the free space as shown in Fig. 6.

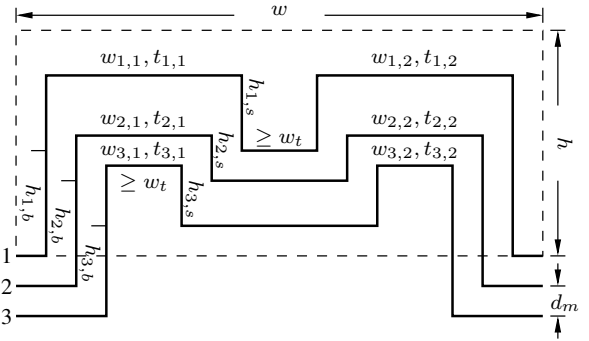

Fig. 6: Model of meander segments for a wire group.

As discussed before, the 0-1 variable $t_{i,j}$ defines whether a subordinate meander segment can exist in the final routing. Examining the patterns in Fig. 6, we can find that a subordinate meander segment can only grow upwards when its upper neighboring wire has such a pattern. Assume that $t_{i,j} = 1$ when a subordinate meander segment exists. The vertical dependency constraint can be modeled as

$$t_{i,j} \geq t_{i+1,j}, i = 1, \ldots M - 1. \quad (4)$$

According to this constraint, if a subordinate meander segment does not exist, the ones below it can not be created to extend wire lengths. That is to say, all the subordinate meander segments below it are blocked due to the nature of the chained constraint (4). In Fig. 6 the widths of the meander segments should be no smaller than the given minimum distance $w_t$. Therefore, we can model the constraints for the width $w_{i,j}$ of the $j$th subordinate meander segment on the $i$th wire as

$$w_{i,j} \geq t_{i,j}w_t + \sum_{k=i+1}^{M} 2t_{k,j}d_m \quad (5)$$

where $d_m$ is the minimum space between wires and $\sum_{k=i+1}^{M} 2t_{k,j}d_m$ is the sum of increased widths due to the meander segments which are surrounded below the $i$th wire. From Fig. 6 we can observe that the width of the complete pattern is bounded by the width of the first wire. This width must be no larger than the width of the given free space, so that we have

$$\sum_{k=1}^{N} w_{1,k} + \sum_{k=2}^{N} t_{1,k}w_t \leq w \quad (6)$$

where $w$ is the width of the given free space shown in Fig. 6, and $\sum_{k=1}^{N} w_{1,k}$ is the sum of the widths of the subordinate meander segments on the first wire as constrained in (5). $\sum_{k=2}^{N} t_{1,k} w_t$ is the sum of the distances between the subordinate meander segments on the first wire. Each of these distances should be no smaller than the minimum width $w_t$ of meander segments, as shown by $\geq w_t$ on the first wire in Fig. 6.

With the 0-1 variables $t_{i,j}$, we provide more freedom for the solver to select how many meander segments should be created in the free space and how many wires should be pushed together. Consider the cases illustrated in Fig. 3. If wires tend to be pushed into the free space as a group, the number of meander segments becomes smaller than in the case that only a few wires are pushed into the free space, as in Fig. 3(a). Here exists a tradeoff between sharing the free space among a group of wires and maximizing the length compensation of individual wires. When we model the possible patterns in a free space, we do not know how many meander segments should be created to achieve an optimal solution. But this number has an upper bound, which can be computed as the number of meander segments when only one wire is pushed, similar to the case in Fig. 3(a). With this analysis, we compute the maximum number of possible meander segments in a free space as

$$N = \lfloor (w - w_t)/2w_t \rfloor + 1 \tag{7}$$

where $w$ is the width of the free space and the symbol $\lfloor \ \rfloor$ represents the greatest integer no larger than the parameter. Note that $N$ is the maximum number of possible meander segments. If a group of wires form meander segments together in the free space, that is, they are pushed together into the free space similar to Fig. 3(b), the number of meander segments drops significantly. Consider a wire group with $M$ wires. The maximum number of meander segments for the first wire can be calculated using (7), but the $M$th wire can not have as many meander segments, because creating a meander segment on the $M$th wire requires that all the wires above should be pushed upwards. Therefore, fewer meander segments can be created in the free space. For example, in Fig. 3(b), wire 3 can have at most two meander segments. With this observation, we can reduce the number of 0-1 variables $t_{i,j}$ for different wires. If a meander segment on the $i$th wire in a wire group should be created, the minimum width of the corresponding meander segments on the first wire can be calculated using (5). As defined by (4), if the $i$th wire is pushed upwards, $t_{i,j}$ should be set to 1 and thus $t_{k,j}, k = 1, \ldots i-1$ should also be 1. According to (5), the width of the uppermost meander segment should meet

$$w_{1,j} \geq w_t + \sum_{k=2}^{i} 2d_m = w_1^i. \tag{8}$$

Similar to (7), the maximum number of meander segments $N_i$ for the $i$th wire can be computed as

$$N_i = \left\lfloor (w - w_1^i)/(w_1^i + w_t) \right\rfloor + 1. \tag{9}$$

Comparing (7) and (9), we can find that $N_i$ may be much smaller than $N$ for the $i$th wire. To reduce the number of variables, we set the last $N - N_i$ 0-1 variables for the $i$th wire to zero, as $t_{i,j} = 0, j = N_i + 1, \ldots N$, because these meander segments will never be feasible.

With the constraint (9), we force the solver to grow meander segments at the first $N_i$ positions for the $i$th wire. In order to align the meander segments on different wires so that they can be formed into a group as in Fig. 6, we add additional constraints for all the wires as

$$t_{i,j} \geq t_{i,j+1}, i = 1, \ldots M, j = 1, \ldots N - 1. \tag{10}$$

These constraints force the solver to select the meander segments from the beginnings of the wires. If a meander segment is not selected, all the following ones on the same wire can not be selected either. Note that adding these constraints does not affect the compensated wire lengths, because the new constraints simply rearrange the order of the freely selected meander segments.

In forming meander segments in free spaces, we can increase the lengths of wires to compensate the meander segments removed from the original routing as described in Section III-A. For the $i$th wire in Fig. 6, the compensated length can be expressed as

$$l_i = l_{i,b} + \sum_{k=1}^{N} 2t_{i,k} h_{i,s} \tag{11}$$

where $\sum_{k=1}^{N} 2t_{i,k} h_{i,s}$ is the sum of the heights of all the subordinate meander segments which appear in the final routing with $t_{i,k} = 1$. $l_{i,b}$ is newly defined here to represent the wire length increased by the height of the base meander segment as discussed using Fig.5 before. If there exist any subordinate meander segments on the $i$th wire, the base meander segment is always included in the model to shift space to the wires below, so that $l_{i,b} = 2h_{i,b}$ as illustrated in Fig. 6 too; otherwise $l_{i,b}$ is equal to zero. This description is equivalent to

$$\text{if } \sum_{k=1}^{N} t_{i,k} \geq 1, \text{ then } l_{i,b} = 2h_{i,b}; \text{ else } l_{i,b} = 0. \tag{12}$$

Considering the constraints (10) we can find that the condition $\sum_{k=1}^{N} t_{i,k} \geq 1$ is equivalent to $t_{i,1} = 1$, because any $t_{i,j} = 1, j = 2, \ldots N$ requires that $t_{i,1} = 1$. Therefore, we can transform the constraint in (12) to

$$\text{if } t_{i,1} = 1, \text{ then } l_{i,b} = 2h_{i,b}; \text{ else } l_{i,b} = 0 \tag{13}$$

$$\iff l_{i,b} = 2t_{i,1} h_{i,b}. \tag{14}$$

Using (11) and (14), we can express the increased length of a wire using the sum of products of a 0-1 variable with either the base height $h_{i,b}$ or subordinate height $h_{i,s}$. This formulation is in a quadratic form because it contains the sum of $t_{i,k} h_{i,s}$ and $t_{i,1} h_{i,b}$ where all the values of $t_{i,k}$, $h_{i,s}$, $t_{i,1}$ and $h_{i,b}$ should be determined by the solver. We will explain a method to transform this formulation into an ILP formulation later.

The model in Fig. 6 describes the pattern in one free space. However, it is very common that a group of wires has free spaces at both sides of it. In this case, some wires may form meander segments in the upper space and others may use the lower space, as illustrated in Fig. 7. As we have discussed before, if more wires are pushed into a free space, fewer meander segments can be formed. Therefore, we should allow the solver to select the direction of the meander segments on each wire. To meet this requirement, we assign a new 0-1 variable $t_i$ for the $i$th wire. If the meander segments on this wire are upward, $t_i$ is set to 1; otherwise, $t_i$ is set to 0. Because a wire can be pushed upwards only when the wires above it are pushed upwards, we can establish the relation between the new variables as

$$t_i \geq t_{i+1}, i = 1, \ldots, M-1 \tag{15}$$

which also implies the downward constraints

$$1 - t_i \leq 1 - t_{i+1}, i = 1, \ldots, M-1. \tag{16}$$

Now consider the patterns in Fig. 6. If the solver determines the $i$th wire should form upward patterns, this wire must be allowed to be pushed upwards, which means

$$t_{i,j} \leq t_i, j = 1, \ldots, N. \tag{17}$$

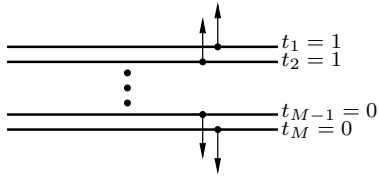

Fig. 7: Selecting directions of meander segments.

If the $i$th wire is pushed downwards, the patterns in Fig. 6 are flipped and the constraints similar to (17) should be written as

$$t_{i,j} \leq 1 - t_i, \, j = 1, \ldots, N. \tag{18}$$

In finding an optimal solution, the solver can determine which wires should be pushed upwards or downwards. That is, the wires are partitioned to two groups automatically by setting the values $t_i, i = 1, \ldots, M$, so that a proper number of meander segments $N_i$ as defined in (9) for each wire can be chosen to establish a balanced length compensation on all wires. Note here we assign only one variable $t_i$ for the $i$th wire, so that all the meander segments on a wire should be pushed upwards or downwards at the same time. We do not allow individual selection of the direction of each meander segment, so that the number of new variables can be reduced. If we find that some wires can not be compensated very well in the iterations, we shift all the wire groups downwards or upwards to merge the two free spaces, so that both areas can be used by the wires for a better compensation.

*2) Modeling the sharing free space by multiple wire groups*

In the last section, we have explained how to model the meander patterns in a free space. After the original dense meander segments are removed, there may be some free spaces surrounded by different wire groups. All these wire groups may grow meander segments into the same space, leading to a resource sharing problem. Fig. 8 shows an example of space sharing. In this example, the free space $S$ represented by the rectangular area is shared by the wire groups around $S$. For a wire group $wg_i$, a free space $S_i$ is allocated from $S$ to grow meander segments. But the dimensions of $S_i$ should be determined considering the relations of other wires and required length compensation globally. In this paper, we use a model also based on the 0-1 selection variables to describe the possible combinations of allocated free spaces for wire groups, and an ILP solver is used to find out an optimal solution.

We use the wire group $wg_i$ in Fig. 8 as an example to explain the model for space sharing. From $wg_i$ upwards, the free space $S_i$ may have conflicts with the free spaces growing in the horizontal direction. For example, $S_j$ blocks $S_i$ in Fig. 8(a). In Fig. 8(b), $S_i$ blocks $S_j$ but is blocked by $S_k$. To model these conflict conditions, we assign a 0-1 variable $c_{i,j}$ for the wire group pair $(wg_i, wg_j)$. If $c_{i,j} = 1$, $wg_i$ can pass $wg_j$; otherwise, $wg_i$ is blocked by $wg_j$. For example, $c_{i,j} = 0$ in Fig. 8(a), and $c_{i,j} = 1, c_{i,k} = 0$ in Fig. 8(b).

For the wire group $wg_i$, we represent its wire width using $w_i^c$, which is a constant and set after the original dense meander segments are removed. For the free space $S_i$, we describe its horizontal dimension as $x_i$ and its vertical dimension as $y_i$, as shown in Fig. 8(a). Note that $x_i$ and $y_i$ both are variables and should be set by the solver. Because the wires can only be pushed in their perpendicular direction, the dimension $x_i$ of the free space $S_i$ should meet

$$x_i \leq w_i^c. \tag{19}$$

Similarly, a constraint for a wire group $wg_j$ at the left side of the free space $S$ can be written as

$$y_j \leq w_j^c. \tag{20}$$

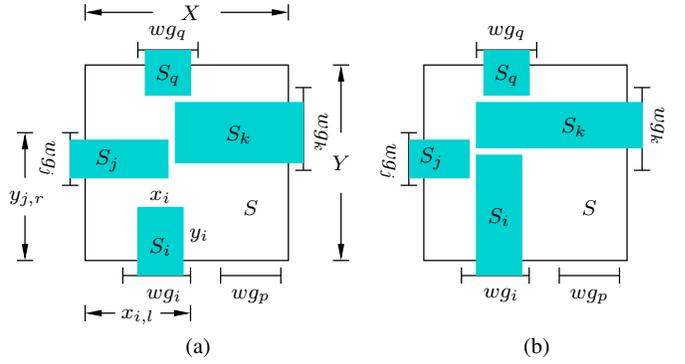

Fig. 8: Model of sharing free space by multiple wire groups.

Because the dimensions of the free spaces for the wire groups, for example, $x_i$ and $x_j$, are all variables, the solver has the freedom to balance the area usage between different wire groups. For example, in Fig. 8(a) the vertical dimension of the free space for $wg_j$ can be set relatively small, so that the free space for $wg_i$ can be enlarged for more length compensation.

The next set of constraints are from the overlapping wire groups at either the horizontal direction or the vertical direction. For example, in Fig. 8 $wg_i$ and $wg_q$ overlap horizontally, and $wg_j$ and $wg_k$ overlap vertically. If free spaces are established from both wire groups in an overlapping pair, the corresponding dimensions should be constrained so that the total dimension of the free space $S$ is not exceeded. For example, the overlapping group pairs $(wg_i, wg_q)$ and $(wg_j, wg_k)$ have the dimensional constraints

$$x_j + d_m + x_k \leq X \tag{21}$$
$$y_i + d_m + y_q \leq Y. \tag{22}$$

Now we establish the constraints from the blocking combinations of wire groups. Using wire group $wg_i$ as an example, we search upwards from it to establish the blocking constraints with other wire groups at the left or the right side of $S$. At first we can find that if $wg_i$ passes a wire group $wg_k$, it should pass the wire group $wg_j$ which is below $wg_k$. If we order the wire groups at the left and right sides of $S$ according to their positions from bottom to top, we can deduce the dependency between the 0-1 variables assigned before as

$$c_{i,j} \leq c_{i,j-1} \tag{23}$$

where the $(j-1)$th wire group is below the $j$th wire group.

In Fig. 8(a), if $wg_i$ is blocked by the wire group $wg_j$ from the left side, that is, $wg_j$ spans above $wg_i$, the vertical dimensions of their free spaces $S_i$ and $S_j$ should meet

$$y_i + d_m + y_j \leq y_{j,r} \tag{24}$$

where $y_i$ and $y_j$ are the dimensions of $S_i$ and $S_j$ in the vertical direction, respectively. $d_m$ is the minimum space between wires. $y_{j,r}$ is the distance from the bottom of $S$ to the left end of $wg_j$. This constraint is only valid when the 0-1 variable $c_{i,j}$ is 0, and we can incorporate this additional constraint into (24) as

$$y_i + d_m + y_j + (1 - c_{i,j})\Gamma \leq y_{j,r} + \Gamma \tag{25}$$

where $\Gamma$ is a predefined very large constant. If $c_{i,j}$ is equal to 0, the constraint is the same as (24). If $c_{i,j}$ is equal to 1, the constraint becomes

$$y_i + d_m + y_j \leq y_{j,r} + \Gamma \tag{26}$$

which always holds for a very large $\Gamma$. With this technique, we can now express the other constraints in the space sharing. The constraint

(25) is valid when $wg_i$ is blocked by $wg_j$. If $wg_i$ can pass $wg_j$, as shown in Fig. 8(b), the width relation should be established as

$$x_j + d_m + x_i + c_{i,j}\Gamma \leq x_{i,l} + \Gamma \quad (27)$$

where $x_{i,l}$ is the distance from the left side of $S$ to the right end of $wg_i$. This constraint is only relevant when $c_{i,j} = 1$; otherwise it always holds and has no effect on the model. If we search from $wg_i$ upwards further, $wg_i$ may block all the wire groups from left and right sides. In this case the sum of the corresponding vertical dimensions of $wg_i$ and the wire groups at the top which $wg_i$ overlaps should not exceed the height $Y$ of $S$, as already defined in in (22).

The constraints (21)–(27) are created for the wire group $wg_i$ at the bottom side. For the wire groups at the top of $S$, for example, $wg_q$, we need to establish their conflict constraints similar to (21)–(27) downwards. Note here we do not establish the relations of the 0-1 variables from a wire group at the top and from a wire group at the bottom. If the solver allows both wire groups to block the same wire groups from left or right sides, some area overlap between the upper group and the lower group seems to appear. However, the constraints (21)–(22) guarantee that there is no such overlap in the final allocation of free spaces for wire groups, and the solver will select an optimal relation between their dimensions to maximize the length compensation.

The horizontal and vertical dimensions in Fig. 8 describe the available free spaces allocated to wire groups. For example, $x_i$ and $y_i$ are the width $w$ and height $h$ in (6) and (3), respectively. With this connection, the relation between the allocation of free spaces for wire groups and the number of possible meander segments as well as the reintroduced wire lengths in Fig. 6 is established and the solver has the freedom to select an optimal solution from all configurations.

Till now we have discussed the constraints for wire group $wg_i$. We repeat this process for all wire groups at the top and the bottom of $S$ to establish all 0-1 variables modeling area conflicts and the corresponding constraints. The constraints of the wire groups from the left side and the right side of $S$ are also handled during this process, for example (21) and (27) so that these wire groups need not to be processed separately. The constraints for space sharing in this section and the constraints modeling patterns of meander segments in Section III-B1 form an optimization problem, and will be solved in Section III-C.

In real routing, the free space after removing the original meander segments may be irregular. When identifying these spaces, we try to determine the largest rectangular area for each wire group. For example, in Fig. 8 the wire groups $wg_i$ and $wg_p$ may have different available heights $Y$ because they may not be aligned in the original routing. This identification process, however, still leaves some free spaces unused. In the next section, we will introduce an iterative algorithm to improve the efficiency of space usage.

### C. Solving the model and an iterative algorithm

Till now we have discussed the constraints in an available free space and the sharing of free spaces between different wire groups. In this section, we will formulate the ILP problem and explain an iterative algorithm to compensate the length of meander segments.

In Section III-B1, we have assumed that we know the minimum required length of meander segments $w_t$. However, in improving a given routing, we aim to achieve the largest possible width of meander segments. In the proposed method, we apply a binary search to find the largest feasible $w_t$. In each iteration of the binary search, the model in Section III-B1 and III-B2 is established and a solver is applied to determine the values of the variables in the constraints.

In each iteration with a given minimum width of meander segments, we formulate an optimization problem in what follows. From Fig. 6, we can see that the compensated wire lengths come from the heights of newly created meander segments. For a free space, the sum of compensated lengths of a wire is defined in (11). Assume that we have in total $M_t$ wires in the design and the $j$th wire belongs to $n_j$ wire groups in creating the patterns in free spaces. We write the indexes of this wire in the $n_j$ groups as $i_1, i_2, \ldots i_{n_j}$. Therefore, the total compensated length for the $j$th wire can be computed using (11), as

$$L_j = \sum_{k=1}^{n_j} l_{i_k}. \quad (28)$$

After removing the dense meander segments from the original routing, we know how much wire length we should compensate so that the improved routing can maintain the same length for each wire as in the original routing. Such a length for the $j$th wire is a constant and represented by $L_j^c$. To guarantee the same wire lengths, we need the following constraints,

$$L_j = L_j^c, j = 1, \ldots M_t. \quad (29)$$

Therefore, the goal of the optimization is to find a solution for all the variables involved in the constraints established in Section III-B1 and III-B2 so that (29) holds. In this optimization problem, the compensated wire lengths contains quadratic terms, for example, $t_{i,k}h_{i,s}$ in (11). In this term $t_{i,h}$ is a 0-1 variable so that we can convert all these quadratic terms to linear form. By changing the index $i$ to $i_k$ and setting $z_{i_k,v} = t_{i_k,v}h_{i_k,s}, z_{i_k,v} \geq 0$, we can transform (11) to

$$l_{i_k} = l_{i_k,b} + \sum_{v=1}^{N} 2z_{i_k,v}. \quad (30)$$

According to [10], the new constraint $z_{i_k,v} = t_{i_k,v}h_{i_k,s}$ can be split into $z_{i_k,v} \geq t_{i_k,v}h_{i_k,s}$ and $z_{i_k,v} \leq t_{i_k,v}h_{i_k,s}$. These split constraints can be transformed into linear forms as

$$z_{i_k,v} \geq t_{i_k,v}h_{i_k,s} \Leftrightarrow z_{i_k,v} \geq h_{i_k,s} - (1 - t_{i_k,v})\Gamma \quad (31)$$
$$z_{i_k,v} \leq t_{i_k,v}h_{i_k,s} \Leftrightarrow z_{i_k,v} \leq h_{i_k,s} \text{ and } z_{i_k,v} \leq t_{i_k,v}\Gamma \quad (32)$$

where $\Gamma$ is a predefined very large constant. These transformations can be verified by checking the equivalence when $t_{i_k,v}$ is set to 0 or 1 individually. Similarly, we can also transform the quadratic terms in (14) and accordingly $l_{i_k,b}$ in (30) to linear forms so that the compensated wire length $L_j$ in (28) is converted into a linear form.

The model explained in Section III-B1 and III-B2, though flexible, still can not cover all the possibilities to grow new meander segments and may leave some free space unused. To improve the efficiency of area usage, we run the modeling and solving process by several iterations, each of which uses the result of the last one as input. With this concept, we try to maximize the compensated wire lengths in each iteration. Instead of using the constraints in (29), with which the solver may report an infeasible solution in the first iteration and produce no result for further iterations, we use the following constraints

$$L_j \leq L_j^c, j = 1, \ldots M_t. \quad (33)$$

Therefore the new optimization problem can be described as

$$\text{maximize: } \sum_{j=1}^{M_t} L_j \quad (34)$$

$$\text{subject to: all } \textit{linear} \text{ constraints in} \quad (35)$$
$$\text{Section III-B1 to III-C except (29).}$$

This formulation is an ILP problem and can be solved directly.

The pseudo code of the complete algorithm is shown in Algorithm 1. The main loop from L9 to L25 implements a binary search

**Algorithm 1:** Alleviation of meander segments using ILP

**Input**: given routing $\mathcal{R}_i$
**Output**: improved routing $\mathcal{R}_o$

- L1 **Parameters:**
- L2 $\mathcal{R}$: a routing with wire lengths to be compensated;
- L3 $\mathcal{M}$: an ILP model from a given routing $\mathcal{R}$;
- L4 $\mathcal{S}$: a solution of the ILP model $\mathcal{M}$;
- L5 $\mathcal{W}$: a set of wires with lengths to be compensated;
- L6 $w_t, \overline{w_t}, \underline{w_t}$: the minimum width of meander segments and its upper and lower bounds.
- L7 $\mathcal{R}_o = \mathcal{R}_i$;
- L8 $w_t = \overline{w_t}$;
- L9 **repeat**
- L10     $\mathcal{R}$=remove_meander_segments($\mathcal{R}_i, w_t$);
- L11     **for** $j=1$ **to** $n$ **do**
- L12         $\mathcal{M}$=create_ILP($\mathcal{R}$);
- L13         $\mathcal{S}$=Solve($\mathcal{M}$);
- L14         $\mathcal{R}$=grow_patterns($\mathcal{R}, \mathcal{S}$);
- L15         $\mathcal{W}$=critical_wires($\mathcal{S}$);
- L16         **if** $\mathcal{W}$ *is empty* **then**
- L17             $\mathcal{R}_o = \mathcal{R}$;
- L18             $\underline{w_t} = w_t$;
- L19             $w_t = (\underline{w_t} + \overline{w_t})/2$;
- L20             go to L25;
- L21         **end**
- L22     **end**
- L23     $\overline{w_t} = w_t$;
- L24     $w_t = (\underline{w_t} + \overline{w_t})/2$;
- L25 **until** $\overline{w_t} - \underline{w_t} < step$;
- L26 **return** $\mathcal{R}_o$;

of the minimum width $w_t$ of meander segments. The inner loop L11 to L22 runs the modeling and ILP solver $n$ times, where $n$ is a predefined number. Each inner iteration improves the result from the last iteration to grow meander segments in the unused areas. If all the lengths are compensated after solving the ILP problem, meaning that no critical wire exists in the result (L16), a feasible routing has been found and the iteration continues to try a larger $w_t$. When the distance between $\underline{w_t}$ and $\overline{w_t}$ is smaller than the predefined $step$, the improved routing $\mathcal{R}_o$ is returned.

## IV. EXPERIMENTAL RESULTS

In this section, we show experimental results by applying the proposed post-processing method on several PCB routings. The proposed framework was implemented using C++. The experiments were performed using a computer with a 2.2 GHz CPU and 8 GB memory. We used three test cases from [4] and one case of the BSG routing which is directly illustrated in [3]. These two methods can route given designs efficiently while considering matching wire lengths. However, many dense meander segments still remain in the results from these methods.

In our experiments, we set very strict area constraints so that newly created meander segments can only use the free spaces between wires. With this setting, the general shapes of the buses in the original routings can be retained, but the upper bound of the minimum width of meander segments $\overline{w_t}$ in Algorithm 1 can rarely exceed 3 due to the limited free space, so that we set $\overline{w_t} = 3$ for most test cases. The routing of case4 from [3] is very tight and actually no much space left between the wires. Therefore, we set $\overline{w_t} = 2$ for this case in the experiments. The minimum widths of the meander segments in these routings are equal to 1, so that $\underline{w_t}$ in Algorithm 1 was set to 1. The ILP solver we used is Gurobi [11].

The results of these test cases are shown in Table I. In the third case from [4] there are six buses. We report the results of meander segment alleviation of these buses individually, because they have different achievable minimum width $w_t$. The meaning of columns in Table I is described as follows.

- $n_w$: the number of wires having length to be compensated.
- $l_{max}$: the maximum length to be compensated on a wire.
- $l_{avg}$: the average length of all wires on a bus to be compensated.
- $l_{total}$: the total length of the bus to be compensated.
- $n_v$: the number of variables.
- $n_c$: the number of constraints.
- $n_{v_p}$: the number of variables after preprocessing by the solver.
- $n_{c_p}$: the number of constraints after preprocessing by the solver.
- $n_{iter}$: the number of all iterations.
- $w_t$: the minimum width of meander segments in the final routing.
- $T(s)$: total runtime in seconds.

In Table I we first compare the numbers of variables and constraints before and after preprocessing by the Gurobi solver to reduce the numbers of variables and constraints. The columns $n_v$ and $n_{v_p}$ before the first iteration show that more than half of the variables are removed from the ILP problem by the pre-solver. The columns $n_c$ and $n_{c_p}$ also show a similar trend. This reduction comes from the fact that the requirement of maximizing the compensated wire length as in (34) can actually fix the assignments of many patterns in free spaces directly. Although the solver can determine this setting mathematically, it is not easy to analyze these relations when creating the model. Actually this is the motivation of using such a mathematical model to find the configuration of creating meander segments in different free spaces, because the proposed model describes the relations between wires and their free spaces and the solver provides a global view for the space assignment.

In the columns after the first iteration, the number of wires $n_w$ with uncompensated lengths decreases drastically compared with the number of wires before the first iteration. The maximum uncompensated length $l_{max}$ on a wire, the average and the total lengths, $l_{avg}$ and $l_{total}$, respectively, also drop significantly. This trend confirms the effectiveness of the model because by only one iteration most of the wires are compensated completely to have the same wire lengths as in the original routing. After the first iteration, the numbers of variables and constraints to the ILP solver are also reduced tremendously due to the smaller number of wires of with uncompensated lengths and fewer free spaces.

In the columns of final results, $n_{iter}$ is the number of all iterations used in Algorithm 1 with $step$ set to 0.25, that is, a fourth of the minimum wire space. The minimum widths of meander segments in the final routings are reported as $w_t$. With this minimum width of meander segments, the wires in the final routing have the same wire lengths as in the original one. From $w_t$ we can observe that for most buses the proposed method can increase the minimum width of meander segments to more than two times, and thus reduce the speedup effect effectively. For the sixth bus in case3, we can only achieve the same minimum width as in the original routing, because these wires are tightly bounded and not enough space exists for extending the wire distances. We illustrate the refined routing of case4 from [3] in Fig. 9, where we can see that the meander segments are distributed along the wires evenly, and $w_t$ is equal to 1.5 times of the original width of meander segments in this case. The runtimes of the proposed method are reported as $T$ in seconds, from which we can observe that the proposed method has a high efficiency on all cases except case4. As shown in Fig. 9(a), case4 contains very tight dense meander segments and there is not enough room for detouring wires which interleave in complex forms. In addition, the free spaces after removing dense meander segments are extremely large so that the number of possible combinations for space sharing is also tremendous. Therefore, finding an optimal compensation configuration in this case is very time-consuming. In our future work, we will try to improve the efficiency of the proposed method in solving cases of this type, for example, by partitioning wires into small groups and assigning the shared areas between these groups using the space sharing model described in Section III-B2.

TABLE I: Results of routing refinement

| | Before the first iteration | | | | | | | | After the first iteration | | | | | | Final results | | |
|---|---|---|---|---|---|---|---|---|---|---|---|---|---|---|---|---|---|
| | $n_w$ | $l_{max}$ | $l_{avg}$ | $l_{total}$ | $n_v$ | $n_c$ | $n_{v_p}$ | $n_{c_p}$ | $n_w$ | $l_{max}$ | $l_{avg}$ | $l_{total}$ | $n_{v_p}$ | $n_{c_p}$ | $n_{iter}$ | $w_t$ | $T(s)$ |
| case1 | 16 | 280 | 72.82 | 1238 | 4764 | 10663 | 2836 | 5460 | 3 | 33 | 2.44 | 41.5 | 390 | 682 | 17 | 2 | 106.94 |
| case2 | 13 | 72 | 35 | 490 | 1438 | 2735 | 759 | 1312 | 2 | 28 | 2.86 | 40 | 104 | 165 | 18 | 2.25 | 17.66 |
| case3.1 | 16 | 108 | 43.63 | 698 | 2786 | 5682 | 1802 | 3214 | 2 | 10 | 0.75 | 12 | 91 | 139 | 16 | 2.5 | 153.55 |
| case3.2 | 10 | 20 | 10.55 | 116 | 994 | 2280 | 435 | 849 | 0 | 0 | 0 | 0 | 0 | 0 | 12 | 2 | 13.049 |
| case3.3 | 19 | 98 | 28.9 | 578 | 2973 | 6315 | 1433 | 2535 | 3 | 6 | 0.6 | 12 | 226 | 376 | 16 | 2 | 55.692 |
| case3.4 | 16 | 80 | 26.625 | 426 | 2142 | 4634 | 1312 | 2394 | 3 | 35.5 | 2.28 | 36.5 | 153 | 198 | 19 | 2.25 | 66.52 |
| case3.5 | 15 | 94 | 29.625 | 474 | 3309 | 7134 | 1493 | 2654 | 2 | 2 | 0.16 | 2.5 | 36 | 50 | 16 | 1.5 | 74.23 |
| case3.6 | 12 | 28 | 11.33 | 136 | 1853 | 3292 | 1199 | 2040 | 0 | 0 | 0 | 0 | 0 | 0 | 15 | 1 | 25.4 |
| case4 | 35 | 436 | 290.11 | 10444 | 10266 | 24352 | 6768 | 13209 | 14 | 93.5 | 7.11 | 256 | 4786 | 8022 | 12 | 1.5 | 1510.58 |

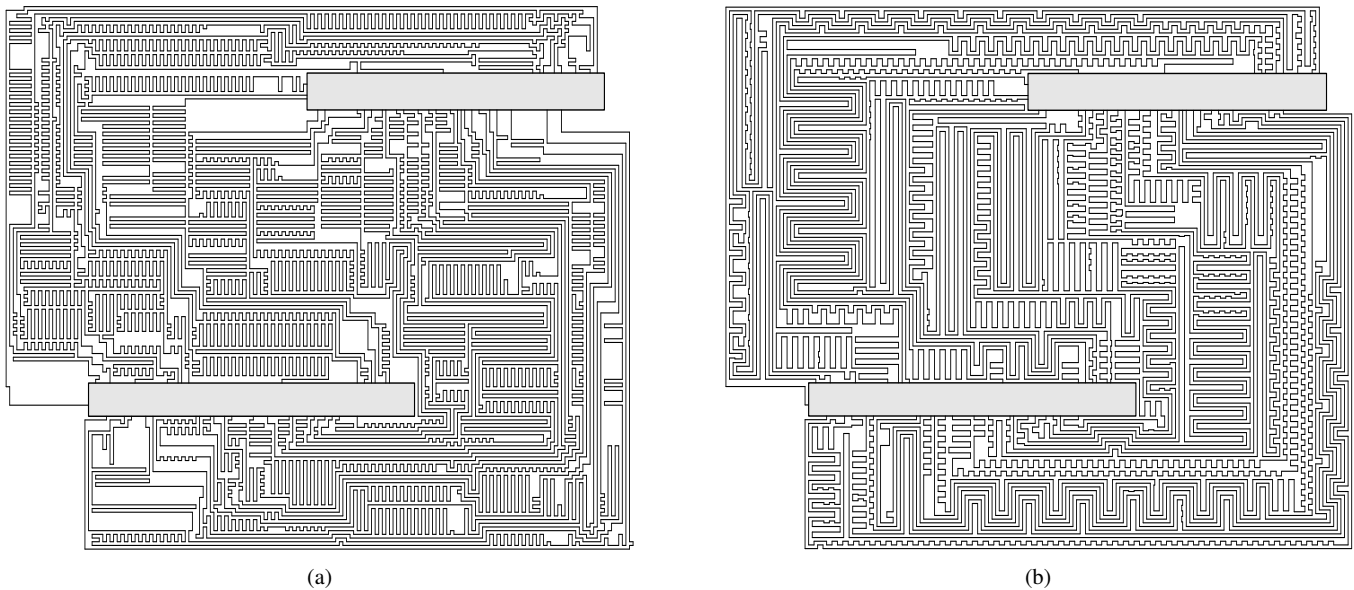

Fig. 9: Routing comparison. (a) Original routing. (b) Refined routing.

## V. CONCLUSION

In this paper we have addressed the delay speedup problem caused by dense meander segments in high-performance PCBs. To extend the widths of these dense segments we proposed a post-processing framework modeling patterns in free spaces and area sharing using 0-1 variables. The problem was transformed into an ILP problem and processed by a solver to provide global balance between uncompensated wire lengths and available free spaces. Experimental results confirm that the proposed method can effectively extend the minimum width of meander segments even to more than two times in most cases, thus reducing the speedup effect significantly.

## ACKNOWLEDGMENT

The work of T.-Y. Ho was supported in part by the Taiwan National Science Council under Grants NSC 101-2220-E-006-016 and NSC 101-2628-E-006-018-MY3, the Ministry of Education, Taiwan, under the NCKU Aim for the Top University Project Promoting Academic Excellence and Developing World Class Research Centers, and the Alexander von Humboldt Foundation.